\documentclass[aps,twocolumn,showpacs,amsmath,amssymb,prl]{revtex4-1}

\usepackage{graphicx}
\usepackage{amsmath}
\usepackage{color}

\newcommand{\be}{\begin{equation}}
\newcommand{\ee}{\end{equation}}
\newcommand{\bfig}{\begin{figure}}
\newcommand{\efig}{\end{figure}}

\begin{document}
\title{Evidence for Unfolded Fermi Surfaces in the Charge-Density-Wave State of Kagome Metal FeGe Revealed by de Haas-van Alphen Effect}

\author{Kaixin Tang$^1$, Hanjing Zhou$^2$, Houpu Li$^3$, Senyang Pan$^{3,4}$, Xueliang Wu$^5$, Hongyu Li$^1$, Nan Zhang$^3$, Chuanying Xi$^4$, Jinglei Zhang$^4$, Aifeng Wang$^5$}
\author{Xiangang Wan$^2$}
 \email{xgwan@nju.edu.cn}
\author{Ziji Xiang$^{1}$}
 \email{zijixiang@ustc.edu.cn}
\author{Xianhui Chen$^{1,3}$}
 \email{chenxh@ustc.edu.cn}

\affiliation{
$^1$ Hefei National Research Center for Physical Sciences at the Microscale, University of Science and Technology of China, Hefei, Anhui 230026, China\\
$^2$ National Laboratory of Solid State Microstructures and School of Physics, Nanjing University, Nanjing 210093, China\\
$^3$ Department of Physics, University of Science and Technology of China, Hefei, Anhui 230026, China\\
$^4$ High Magnetic Field Laboratory, Chinese Academy of Sciences, Hefei, Anhui 230031, China\\
$^5$ Low Temperature Physics Laboratory, College of Physics and Center of Quantum Materials and Devices, Chongqing University, Chongqing 401331, China.
}

\date{\today}
\begin{abstract}
 The antiferromagnetic kagome lattice compound FeGe has been revealed to host an emergent charge-density-wave (CDW) state which manifests complex interplay between the spin, charge and lattice degrees of freedom. Here, we present a comprehensive study of the de Haas-van Alphen effect by measuring torque magnetometry under magnetic fields up to 45.2\,T to map Fermi surfaces in this unusual CDW state. For field along the $c$ direction, we resolve four cyclotron orbits; the largest one roughly corresponding to the area of the 2$\times$2 folded Brillouin zone. Three smaller orbits are characterized by light effective cyclotron masses range from (0.18-0.30)$m_e$. Angle-resolved measurements identify one Fermi surface segment with weak anisotropy. Combined with band structure calculations, our results suggest that features of unfolded Fermi surfaces are robust against CDW reconstruction, corroborating the novel effect of a short-ranged CDW on the electronic structure.
\end{abstract}
\maketitle                   
Kagome-lattice metals possess a unique electronic band structure that hosts Dirac points, van Hove singularities (vHSs) and flat bands; due to the geometrically frustrated interactions and complex sublattice interference in kagome lattice, a plethora of novel quantum phases of matter can arise from these electronic states \cite{Wen_SC,Jianxin_SDW,ThomaleInterference,ThomaleOrders,Qianghua}. Recent experimental achievements, including the realization of distinct topological phases in kagome magnets \cite{Mn3Sn,Fe3Sn2,Co3Sn2S2,TbMn6Sn6} and the observation of unusual density-wave orders \cite{CsV3Sb5muonSR,CsV3Sb5PDW} together with a Potts-type electronic nematicity \cite{NMRnematic,STMnematic,Zeljkovic} in kagome superconductors $A$V$_3$Sb$_5$ ($A$ = K, Rb, Cs), further highlight an abundance of intriguing physical phenomena in kagome lattice . Most notably, unconventional electronic orders emerging from kagome lattices frequently invoke intertwinements between the charge, spin, orbital and lattice degrees of freedom \cite{CsV3Sb5stripe,chiralFS}. Such intertwining orders serve as key ingredients for understanding the intricate kagome physics.

Hexagonal FeGe (space group P6/mmm, No.\,191) has recently been revealed as an ideal platform for exploring the interplays between electronic orders in charge and spin channels \cite{FeGeCDW_Neutron,FeGeCDW_STM,FeGeCDW_ARPES}. This material crystallizes in a B35 layered structure composed of alternatively stacking Fe$_3$Ge kagome planes and Ge$_2$ honeycomb planes \cite{Neutron1966} [inset of Fig.\,1(a)]. At $T_N \simeq$ 410\,K, it develops a collinear A-type antiferromagnetic (AFM) order with Fe spins aligning along the crystallographic $c$-axis ([0001]) \cite{Neutron1966,FGmagnetism1972}, which evolves into a double-cone configuration below $\sim$ 60\,K \cite{Neutron1984}. When a magnetic field ($H$) is applied along $c$, a spin-flop transition occurs at $H_{\rm SF} \simeq$ 7\,T at low temperatures ($T$), above which the Fe spins are aligned predominantly within the kagome plane \cite{FGmagnetism1972,SpinFlop1970}. Latest neutron diffraction, scanning tunneling microscopy (STM) and angle-resolved photoemission spectroscopy (ARPES) experiments \cite{FeGeCDW_Neutron,FeGeCDW_STM,FeGeCDW_ARPES} identify the emergence of a charge-density-wave (CDW) order inside the AFM phase region at $T_{\rm CDW} \simeq$ 100\,K; the CDW order, manifested by a 2$\times$2$\times$2 superlattice, is intimately coupled to the ordered Fe magnetic moment. As pointed out by up-to-date theoretical works \cite{GuoqingChang,WanDM,WanNesting,YilinWang,QimiaoSi}, this CDW order is not caused by conventional electron-phonon coupling; instead, it acts as an rare example of electron-correlation-driven charge order which is strongly intertwined with magnetism. Nonetheless, the exact mechanism responsible for the CDW formation is still under debate.

For a deeper understanding of the nature of unusual CDW in kagome metal FeGe, information about the electronic structure with high momentum resolution is urgently needed. Quantum oscillation study is an unreplaceable tool for this purpose since it can map the Fermi surface (FS) morphology with extreme precision. In this paper, we provide the first report on the observation of de Haas-van Alphen (dHvA) effect ({\it i.e.}, quantum oscillations in magnetization) in the CDW-ordered state of FeGe. Four dHvA frequencies are resolved with $H \parallel c$; the highest one corresponds to an orbit that roughly matches the area of the folded Brillouin zone (BZ) for the 2$\times$2 superlattice ({\it i.e.}, 1/4 of the unfolded BZ). The other three frequencies are within the range of 440-550\,T, indicative of small cyclotron orbits. The strongest dHvA frequency shows a weak dependence on the tilt angle $\theta$ between $\textbf{\textit{H}}$ and $c$ up to $\theta \sim$ 70$^\circ$, highlighting three-dimensional (3D) character of the underlying FS. Moreover, the dHvA frequencies can be reasonably assigned to the unreconstructed FS structure in the AFM state. The band folding effect introduced by CDW is thus verified to be weak. Such a scenario is in accord with the electronic origin and short-ranged nature of CDW in FeGe.

\begin{figure}[htbp!]
\centering
\includegraphics[width=0.40\textwidth]{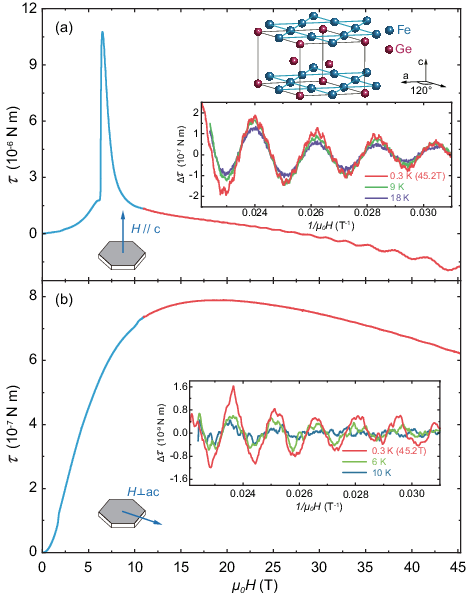}
\caption{Magnetic-field($H$)-dependent magnetic torque $\tau$ measured on a hexagonal FeGe single crystal up to $\mu_0H$ = 45.2\,T; $H$ was applied approximately parallel to the crystallographic (a) $c$-axis ([0001]) and (b) in-plane mirror axis ([01$\bar{1}$0]), respectively. Data were taken at $T$ = 0.34\,K. Red and blue colors represent two separated field-ramping steps for the hybrid magnet (See Sec. I in Supplemental Material \cite{SM}). Insets present the oscillatory torque $\Delta \tau$, plotted against the inverse magnetic field $1/\mu_0H$. $\Delta \tau$ is obtained by subtracting a polynomial nonoscillatory part from $\tau(H)$. Upper inset of (a) shows a schematic illustration of the crystal structure.
}
\label{FIG. 1}
\end{figure}

Single crystals of hexagonal FeGe were synthesized using the chemical vapor transport method \cite{FeGeCDW_Neutron}. Most of the dHvA measurements were performed on samples from batch ''A" grown in Hefei. (For details of sample preparation and characterization, see Supplemental Material \cite{SM}.) Torque magnetometry data were measured using a capacitive magnetometer at the Chinese High Magnetic Field Laboratory in Hefei. Figures 1(a) and (b) display the magnetic torque curves for a FeGe single crystal measured in a $^3$He cryostat with $H \parallel c$ and $H \perp ac$ ({\it i.e.}, along the in-plane mirror axis [01$\bar{1}$0]), respectively, up to a record-breaking static magnetic field of 45.2\,T supplied by a hybrid magnet. For $H \parallel c$, a sharp anomaly appears at $\mu_0H_{\rm SF}$ = 6.5\,T, consistent with the field-induced spin-flop transition \cite{FGmagnetism1972,SpinFlop1970}. Clear dHvA oscillations develop in the spin-flop state. with a pattern containing both low-frequency ($F$) and high-$F$ components [inset of Fig.\,1(a)]. On the torque curve measured with $H \perp ac$, the spin-flop transition is absent, and weak dHvA wiggles emerge only above $\sim$ 30\,T [inset of Fig.\,1(b)].

The oscillatory magnetic torque [$\Delta\tau$, insets of Figs.\,1(a) and (b)] incorporates all the crucial information for a dHvA study. Within the framework of 3D Lifshitz-Kosevich (LK) theory \cite{Shoenberg},
\begin{equation}
 \begin{split}
 \Delta\tau =& A_0 B^{3/2}\sum_{i} \frac{dF_i}{d\theta} |\frac{\partial^2 S_{i,k}}{\partial k_{\parallel}^{2}}|^{-1/2}
 \sum_{p=1}^{\infty} p^{-3/2} R_T R_D R_S \\
 &sin[2\pi p(\frac{F_i}{B}+\frac{\Phi_{i}}{2\pi})\pm\frac{\pi}{4}].
 \end{split}
 \label{dHvA}
\end{equation}
Here $A_0$ is a constant, the two sums run over band indices $i$ and oscillation harmonics $p$, respectively; $S_i$ is the extremal momentum-space area of the $i$th FS (according to the Onsager relation, $F_i=\frac{\hbar}{2\pi e}S_i$), $k_{\parallel}$ is the momentum component parallel to $H$, $R_T$, $R_D$ and $R_S$ are the thermal, Dingle and spin-splitting damping factors, respectively, and $\Phi_i$ the total phase of quantum oscillation. Due to the small magnetization $M$ of AFM FeGe (Fig.\,S2 in \cite{SM}), we have the magnetic induction $B \approx \mu_0H$.

\begin{figure}[htbp!]
\centering
\includegraphics[width=0.41\textwidth]{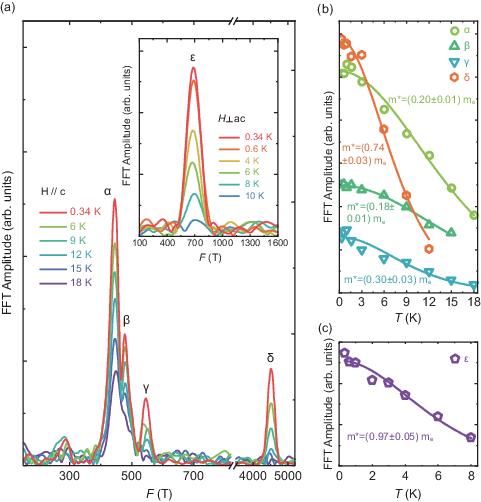}
\caption{(a) Fast Fourier transform (FFT) spectra of the dHvA data for $H \parallel c$ measured at varying temperatures from 0.34 to 18\,K. The FFT windows used for resolving the low-$F$ (below 850\,T) and high-$F$ (above 3500\,T) components are 12-44.65\,T and 32-44.65\,T, respectively; FFTs in the two frequency ranges are shown in different amplitude scales. Inset: the FFT spectra for $H \perp ac$ configuration. (b) and (c) $T$ dependence of FFT peak amplitudes for oscillation frequencies shown in (a). Solid lines are fits to the thermal damping factor $R_T$ in the Lifshitz-Kosevich (LK) formula \cite{Shoenberg}.}
\label{FIG. 2}
\end{figure}

\begin{figure}[htbp!]
	\centering
	\includegraphics[width=0.46\textwidth]{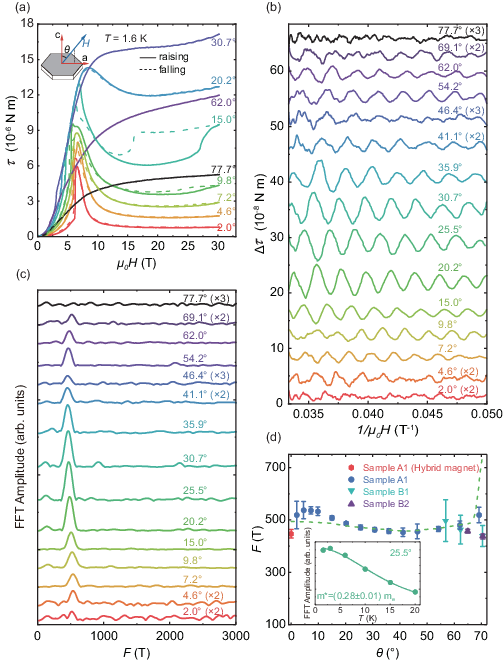}
	\caption{(a) Magnetic torque $\tau(H)$ of FeGe at various angles $\theta$, measured with $\textbf{\textit{H}}$ rotated in the $ac$-plane (inset sketch). Solid and dashed lines are upsweep and downsweep curves, respectively. (b) Oscillatory torque $\Delta\tau$ for different $\theta$. (c) The FFT spectra of the dHvA data shown in (b); the single FFT peak is indexed as $F_{\alpha}$. In (b) and (c), data at several $\theta$ are multiplied by a factor for clearance. (d) Angular dependence of $F_{\alpha}$ measured in various samples. Green dashed line is a simulation given by a hypothetical FS plotted in Fig.\,4(d). Inset: $T$-dependence of dHvA amplitude measured at $\theta$ = 25.5$^\circ$. Solid line is the LK fit.
	}
	\label{Fig3}
\end{figure}

We first look into the dHvA frequencies by performing fast-Fourier transform (FFT) on the $\Delta\tau$(1/$\mu_0H$) curves. As shown in Fig.\,2(a), for $H \parallel c$, four fundamental frequencies can be resolved: $F_{\alpha}$ = 445\,T, $F_{\beta}$ = 478\,T, $F_{\gamma}$ = 546\,T and $F_{\delta}$ = 4515\,T. The FFT peak of $F_{\alpha}$ is the strongest, suggesting that branch $\alpha$ dominates the dHvA spectrum. The observation of the frequency component $F_{\delta}$ is of particular interest, since the corresponding orbit area reaches $\simeq$ 23.6$\%$ of the in-plane first BZ for the unreconstructed unit cell ($\sim$ 19130\,T, taking the lattice constant $a$ = 4.985 \AA \cite{FeGeCDW_STM}). In other words, this extremal orbit has an area roughly equal to the reconstructed BZ in the scenario of 2$\times$2 in-plane CDW modulation. For $H \perp ac$, the dHvA oscillation waveform is characterized by a single frequency, $F_{\epsilon}$ = 687\,T [inset of Fig.\,2(a)]. This branch stems from a closed orbit lying in a plane parallel to the interlayer ($c$) direction. No higher harmonics ($p >$ 1) are discernible for any of these frequencies.

The $T$-dependent amplitudes of dHvA oscillations ({\it i.e.,} heights of FFT peaks) are fitted to the damping factor $R_T$ in the LK model (Eq.\,\ref{dHvA}): $R_T = X/\sinh(X)$, where $X = 14.69\,(m^*/m_e) T/B$, $m^*$ is the effective cyclotron mass and $m_e$ the free electron mass \cite{Shoenberg}. The fitting results are presented in Figs. 2(b) and (c). Light cyclotron masses are revealed for dHvA branches $\alpha$ ($m^* = $ 0.20$\pm$0.01 $m_e$), $\beta$ ($m^* = $ 0.18$\pm$0.01 $m_e$) and $\gamma$ ($m^* = $ 0.30$\pm$0.03 $m_e$). As for the high-$F$ component $\delta$ and the vertically extended cyclotron orbit $\epsilon$, the values of $m^*$ are fitted to be 0.74$\pm$0.03 and 0.97$\pm$0.05\,$m_e$, respectively. Considering the predominant Fe 3$d$ orbital character of conduction electrons in FeGe \cite{GuoqingChang,WanNesting}, the absence of remarkable quasiparticle mass enhancement points towards a weak correlation effect for the detected electronic bands \cite{footnote}. Analysis of the Dingle damping factor $R_D$ in Eq.\,\ref{dHvA} [Fig.\,S4 in \cite{SM}] provides information of the scattering rate, which in turn gives a quasiparticle mean free path $l_{QP}$ = (80$\pm$20)\,nm for main branch $\alpha$ \cite{meanfreepath}.

To investigate the anisotropy of FSs, we mounted the capacitive magnetometer on a rotation stage and measured the angle-resolved magnetic torque in a 31\,T water-cooled Bitter magnet. As $\textbf{\textit{H}}$ is rotated away from the $c$ direction, the anomaly at $H_{\rm SF}$ becomes broadened and is smeared out above $\theta \sim 20^\circ$; with further increasing $\theta$, $\tau$ evolves into a monotonic function of $H$ [Fig.\,3(a), see Fig.\,S5 in \cite{SM} for additional angle-resolved data]. Within a small angular range of a few degrees around $\theta$ = 15$^\circ$, a wide hysteresis loop appears on $\tau(H)$, which may reflect the formation of magnetic domains with different canted spin components. dHvA oscillations on $\tau$ show up till $\theta \simeq$ 70$^\circ$ [Fig.\,3(b)] and are absent for higher angles. We point out that up to $\theta \sim 80^\circ$, the profile of $\tau(H)$ can be satisfyingly fitted to an AFM model with a spin-flop field $H_{\rm SF} \simeq$ 6-8\,T (See Sec. II in Supplemental Material \cite{SM}). Therefore, the dHvA signals are developed in a high-$H$ spin-flop state.

Because of the lower maximum field of the Bitter magnet, only one dHvA frequency can be unequivocally resolved in our rotation experiments [Fig.\,3(c)]. We assign it to the strongest main peak $F_{\alpha}$ [Fig.\,2(a)]. As plotted in Fig.\,3(d), $F_{\alpha}$ has a nonmonotonic but weak angular dependence with its value varying between 440 and 540\,T up to $\theta \sim$70$^\circ$, above which it suddenly disappears. LK fit at $\theta =25.5^{\circ}$ yields $m^*$ = 0.28$\pm$0.01 $m_e$ [inset of Fig.\,3(d)]. These results, together with the observation of orbit $\epsilon$ with $H \perp ac$ [Fig.\,1(b)], imply strong 3D character of FS morphology in the layered compound FeGe.

\begin{figure}[htbp!]
	\centering
	\includegraphics[width=0.44\textwidth]{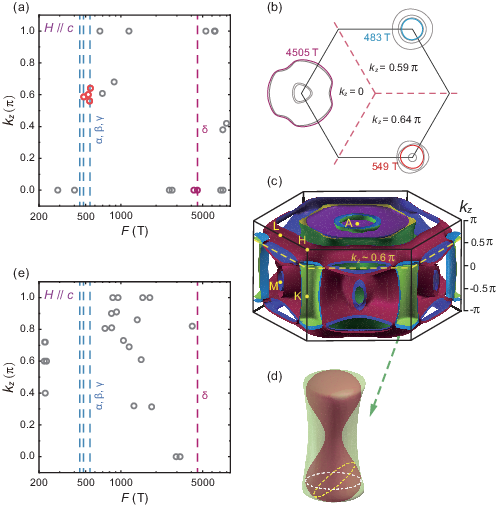}
	\caption{(a) dHvA frequencies for the $H \parallel c$ configuration extracted from the FS maps calculated for the unreconstructed (pristine) structure, plotted against $k_z$. Experimental values are denoted by vertical dashed lines. Red and purple symbols represent the calculated orbit areas that match experimental results, while the gray symbols are absent in our dHvA data. (b) FS contours in the planes of $k_z$ = 0, 0.59\,$\pi$, and 0.64\,$\pi$; colored circles are extremal orbits resolved by dHvA measurements. (c) DFT-calculated FS structure in the spin-flop AFM state for the pristine phase The vertical coordinate axis on the right side shows the value of $k_z$ for the first BZ. (d) An expanded view of the electronic pocket around $K$. The red dumb-bell lobe is an illustration of hypothetical FS pocket which can describe the angle-dependent dHvA frequencies presented in Fig.\,3(d). (e) Same as (a) but calculated for the 2$\times$2$\times$2 CDW-reconstructed phase \cite{FeGeCDW_Neutron} [see Fig.\,S8(d) in \cite{SM}]. }
	\label{FIG. 4}
\end{figure}

By performing first-principles calculations based on the density functional theory (DFT), we obtained the FS structures for both the pristine AFM [Fig.\,S8(b), \cite{SM}] and CDW phases of FeGe (for details, see Sec.\,IV of Supplemental Material \cite{SM}). Extremal orbit areas were extracted and compared with the experimental results. As shown in Fig.\,4(a), all dHvA branches measured with $H \parallel c$ can be accounted for in the unreconstructed band structure calculated for the pristine phase, which is quite unusual for a CDW-ordered material. The dominant $\alpha$ component is assigned to a local maximum ($k_z$ = 0.59 $\pi$) on a FS centered at the $K$ point in BZ, whereas the highest frequency $F_{\delta}$ stems from a large FS sheet that is also around $K$ [Figs.\,4(b) and (c), see also Figs.\,S9 and S10 in \cite{SM}]. $\epsilon$ is likely to be associated with a ``neck" connecting adjacent BZs [Fig.\,S10(d), \cite{SM}]. The weak $\theta$-dependence of $F_{\alpha}$ may indicate a more isotropic segment on the corresponding FS pocket compared with the DFT-calculated one, implying that the real FS pocket is possibly more dumb-bell-shaped [Fig.\,4(d), see also Fig.\,S11, \cite{SM}]. Such deformation with a reduction of $k_F$ in the basal plane potentially reflects the impact of CDW modulation and is indeed consistent with recent ARPES results \cite{JiangARPES}. Nevertheless, DFT calculations based on the 2$\times$2$\times$2 CDW order reported in Ref.\,\cite{FeGeCDW_Neutron} cannot reproduce the observed dHvA frequencies [Fig.\,4(e), see also Fig. S12(b) in \cite{SM}]. We note that additional symmetry limitations have been proposed for the CDW-ordered state \cite{WanDM,YilinWang} and these models are to be tested against our dHvA data.

The absence of CDW-reconstructed orbits can be attributed to two reasons: (i) The weak CDW distortion results in small spectral weights for the folded bands; consequently, the hybridization gap between the intersecting folded and original bands are small, allowing magnetic breakdown recovering the unreconstructed orbits at low $H$ \cite{WilsonTaSe2}. (ii) Due to a small CDW correlation length $\xi$ (compared with $l_{QP}$), electron cannot effectively ``see" the translational symmetry breaking caused by superlattice formation during their cyclotron motion \cite{Harrison_AFM}. It has been pointed out that a very short $\xi$ severely reduces the quantum oscillation amplitude by introducing an additional damping factor $R_{\rm CDW}$, which gives an exponential decay with increasing ratio of the cyclotron radius $r_c$ to $\xi$ \cite{ShortRangeCDW}. Intriguingly, $\xi$ in hexagonal FeGe can be controlled by changing growth and annealing conditions \cite{SM,AnnealingAifengWang}. The dHvA effect is completely absent in samples with a very short $\xi$ of 2-4\,nm \cite{STMYajunYan} (batch ``C", Fig.\,S7 \cite{SM}), verifying the damping effect from $R_{\rm CDW}$ \cite{CDWdamping}. Unexpectedly, in the FeGe samples (batch ``B", grown in Chongqing) with long-range CDW ($\xi >$ 100 nm as revealed by STM \cite{LongRangeCDW}), dHvA oscillations are invisible at low $\theta$ and can only be resolved in a narrow angular range of $\sim55-71^\circ$ (Fig.\,S7, \cite{SM}). Possible explanations include the limitation of $l_{QP}$ from sources other than $\xi$ ({\it e.g.}, point defects), and/or enhanced scattering in the long-ranged CDW state at special $k$-points on the reconstructed orbits that interrupts the cyclotron motion \cite{SM,CDWLinearMR,Tl2201RIXS}.

The dHvA effect in kagome lattice FeGe provide an interesting contrast to the novel quantum oscillations observed in cuprate high-temperature superconductors \cite{YBCO2008Nature,YBCOdHvA,YBCO_XRS,Hg1201Barisic,Hg1201MKChan}, wherein short-ranged CDW is also revealed (with $\xi$ ranging from a few to several tens of nanometers \cite{cuprates}) yet the oscillations originate from CDW-reconstructed FSs. As for FeGe, we propose that the length scales $l_{QP}$, $r_c$ and $\xi$ in our batch ``A" samples establish a unique parameter window in which $R_{\rm CDW}$ is sufficiently large for observing dHvA, yet $\xi$ is not long enough to bring out the reconstructed orbits \cite{LengthEstimation}; both increase and further decrease of $\xi$ appear to be detrimental to the electron orbiting process. Further investigations concerning the underlying microscopic mechanism would grant insights into the nature of electronic system in compounds possessing unusual CDW orders, {\it e.g.}, cuprates, CsV$_3$Sb$_5$ \cite{CsV3Sb5HighF,CsV3Sb5MB} and Ta$_4$Pd$_3$Te$_{16}$ \cite{Ta4Pd3Te16}.

In summary, we unveil the dHvA effect in kagome metal FeGe using torque magnetometry under intense magnetic fields. The dHvA oscillations, observed with both in-plane and out-of-plane field orientations, provide crucial information of the fermiology in the CDW ordered state. The identified cyclotron orbits have light masses and a comparison with the DFT-calculated bands assigns them to the strongly 3D FSs in the {\it unreconstructed} band structure. The folded bands introduced by CDW modulation are not observed; the effect of reconstruction is only reflected by a putative deformation of one initial FS. Such phenomena imply that the emergent charge order in FeGe, which intertwines strongly with magnetism, has a presumably weak and unconventional impact on the FS morphology---probably due to its limited correlation length. Our results add to the abundance of novel physical properties associated with the short-ranged CDW order arising from electron correlations.

\vspace{1ex}

We thank Yilin Wang, Yajun Yan, Ziqiang Wang, Kun Jiang, Rui Wang, Ye Yang, Tao Wu and Zhenyu Wang for helpful discussions. This work was financially supported by the National Key R$\&$D Program of the MOST of China (Grant No. 2022YFA1602602), the National Natural Science Foundation of China (Grants Nos. 12274390, 12188101, 11888101, 12122411 and 12004056), the Anhui Initiative in Quantum Information Technologies (AHY160000) and Chinese Academy of Sciences under contract No. JZHKYPT-2021-08. A.W. acknowledges the Chongqing Research Program of Basic Research and Frontier Technology, China (Grants No. cstc2021jcyj-msxmX0661).


%

%

\begin{thebibliography}{99}

\bibitem{Wen_SC}
W.-H. Ko, P. A. Lee, and X.-G. Wen,
Doped kagome system as exotic superconductor,
Phys. Rev. B {\bf 79,} 214502 (2009).

\bibitem{Jianxin_SDW}
S.-L. Yu and J.-X. Li
Chiral superconducting phase and chiral spin-density-wave phase in a Hubbard model on the kagome lattice,
Phys. Rev. B {\bf 85,} 144402 (2012).

\bibitem{ThomaleInterference}
K. L. Kiesel and R. Thomale,
Sublattice interference in the kagome Hubbard model,
Phys. Rev. B {\bf86}, 121105(R) (2012).

\bibitem{ThomaleOrders}
M. L. Kiesel, C. Platt, and R. Thomale,
Unconventional Fermi surface instabilities in the kagome Hubbard model,
Phys. Rev. Lett. {\bf110}, 126405 (2013).

\bibitem{Qianghua}
W.-S. Wang, Z.-Z. Li, Y.-Y. Xiang, and Q.-H. Wang,
Competing electronic orders on kagome lattices at van Hove filling,
Phys. Rev. B {\bf 87}, 115135 (2013).

\bibitem{Mn3Sn}
K. Kuroda {\it et al.},
Evidence for magnetic Weyl fermions in a correlated metal,
Nat. Mater. {\bf 16}, 1090 (2017).

\bibitem{Fe3Sn2}
L. Ye, M. Kang, J. Liu, F. Von Cube, C. R. Wicker, T. Suzuki, C. Jozwiak, A. Bostwick, E. Rotenberg, D. C. Bell, L. Fu, R. Comin, and J. G. Checkelsky,
Massive Dirac fermions in a ferromagnetic kagome metal,
Nature (London) {\bf555}, 638 (2018).

\bibitem{Co3Sn2S2}
D. F. Liu, A. J. Liang, E. K. Liu, Q. N. Xu, Y. W. Li, C. Chen, D. Pei, W. J. Shi, S. K. Mo, P. Dudin {\it et al.},
Magnetic Weyl semimetal phase in a Kagom\'{e} crystal,
Science {\bf 365,} 1282 (2019).

\bibitem{TbMn6Sn6}
J.-X. Yin {\it et al.},
Quantum-limit Chern topological magnetism in TbMn$_6$Sn$_6$,
Nature (London) {\bf 583,} 533 (2020).

\bibitem{CsV3Sb5muonSR}
C. Mielke III, D. Das, J.-X. Yin, H. Liu, {\it et al.},
Time-reversal symmetry-breaking charge order in a kagome superconductor,
Nature (London) {\bf 602,} 245 (2022).

\bibitem{CsV3Sb5PDW}
H. Chen, H. Yang, B. Hu, {\it et al.},
Roton pair density wave in a strong-coupling kagome superconductor,
Nature (London) {\bf 599,} 222 (2021).

\bibitem{NMRnematic}
L. Nie, K. Sun, W. Ma, {\it et al.},
Charge-density-wave-driven electronic nematicity in a kagome superconductor,
Nature (London) {\bf 604,} 59 (2022).

\bibitem{STMnematic}
P. Wu, Y. Tu, {\it et al.},
Unidirectional electron-phonon coupling in the nematic state of a kagome superconductor,
Nat. Phys. {\bf 19,} 1143 (2023).

\bibitem{Zeljkovic}
H. Li, H. Zhao, B. R. Ortiz, Y. Oey, Z. Wang, S. D. Wilson and I. Zeljkovic,
Unidirectional coherent quasiparticles in the high-temperature rotational symmetry broken phase of $A$V$_3$Sb$_5$ kagome superconductors,
Nat. Phys. {\bf 19,} 637 (2023).

\bibitem{CsV3Sb5stripe}
L. Zheng {\it et al.},
Emergent charge order in pressurized kagome superconductor CsV$_3$Sb$_5$,
Nature (London) {\bf 611,} 682 (2022).

\bibitem{chiralFS}
S. Zhou and Z. Wang,
Chern Fermi pocket, topological pair density wave, and charge-4$e$ and charge-6$e$ superconductivity in kagom\'{e} superconductors,
Nat. Commun. {\bf 13,} 7288 (2022).


\bibitem{FeGeCDW_Neutron}
X. Teng {\it et al.},
Discovery of charge density wave in a kagome lattice antiferromagnet,
Nature (London) {\bf 609,} 490 (2022).

\bibitem{FeGeCDW_STM}
J.-X. Yin, Y.-X. Jiang, X. Teng, M. S. Hossain, S. Mardanya, T.-R. Chang, {\it et al.},
Discovery of charge order and corresponding edge state in kagome magnet FeGe,
Phys. Rev. Lett. {\bf 129,} 166401 (2022).

\bibitem{FeGeCDW_ARPES}
X. Teng, J. S. Oh, {\it et al.},
Magnetism and charge density wave order in kagome FeGe,
Nat. Phys. {\bf 19,} 814 (2023).


\bibitem{Neutron1966}
H. Watanabe and N. Kunitomi,
On the Neutron Diffraction Study of FeGe,
J. Phys. Soc. Jpn. {\bf 21,} 1932 (1966).

\bibitem{FGmagnetism1972}
O. Beckman, K. Carrander, L. Lundgren, and M. Richardson,
Susceptibility measurements and magnetic ordering of hexagonal FeGe,
Phys. Scr. {\bf 6}, 151 (1972).

\bibitem{Neutron1984}
J. Bernhard, B. Lebech, and O. Beckman,
Neutron diffraction studies of the low-temperature magnetic structure of hexagonal FeGe,
J. Phys. F {\bf 14,} 2379 (1984).

\bibitem{SpinFlop1970}
K. Carrander, H. Schwartz and O. Beckman,
Magnetic Phase Diagram of Hexagonal FeGe,
Phys. Scr. {\bf 2,} 313 (1970).

\bibitem{GuoqingChang}
S. Shao, J.-X. Yin, I. Belopolski, J.-Y. You, T. Hou, H. Chen, Y.-X. Jiang, M. S. Hossain, M. Yahyavi, C.-H. Hsu, Y. Feng, A. Bansil, M. Z. Hasan, and G. Chang,
Intertwining of Magnetism and Charge Ordering in Kagome FeGe,
ACS Nano {\bf 17,} 10164 (2023).

\bibitem{WanDM}
H. Zhou, S. Yan, D. Fan, D. Wang, and X. Wan,
Magnetic interactions and possible structural distortion in kagome FeGe from first-principles calculations and symmetry analysis,
Phys. Rev. B {\bf 108,} 035138 (2023).

\bibitem{WanNesting}
L. Wu, Y. Hu, D. Fan, D. Wang, and X. Wan,
Electron-Correlation-Induced Charge Density Wave in FeGe,
Chin. Phys. Lett. {\bf 40,} 117103 (2023).


\bibitem{YilinWang}
Y. Wang,
Enhanced spin-polarization via partial Ge-dimerization as the driving force of the charge density wave in FeGe,
Phys. Rev. Mater. {\bf 7,} 104006 (2023).

\bibitem{QimiaoSi}
C. Setty, C. A. Lane, L. Chen, H. Hu, J.-X. Zhu, and Q. Si,
Electron correlations and charge density wave in the topological kagome metal FeGe,
arXiv: 2203.01930 (2022).



\bibitem{SM}
See Supplemental Material for details of sample growth and characterizations, first-principles calculations and analysis of torque data using an AFM model, which includes Refs.\,\cite{AnnealingAifengWang}-\cite{JiangARPES}.

\bibitem{AnnealingAifengWang}
X. Wu, X. Mi, L. Zhang, X. Zhou, M. He, Y Chai, and A. Wang,
Annealing tunable charge density wave order in a magnetic kagome material FeGe,
arXiv: 2308.01291 (2023).

\bibitem{HuMiao}
H. Miao {\it et al.},
Signature of spin-phonon coupling driven charge density wave in a kagome magnet,
Nat. Commun. {\bf 14,} 6183 (2023).

\bibitem{LongRangeCDW}
Z. Chen {\it et al.},
Long-ranged charge order conspired by magnetism and lattice in an antiferromagnetic Kagome metal,
arXiv: 2307.07990 (2023).

\bibitem{SpinFlop_BEDT}
H. Uozaki, T. Sasaki, S. Endo, and N. Toyota,
Antiferromagnetic Ordering in the Conducting $\pi-d$ System $\kappa$-(BEDT-TSF)$_{2}$FeCl$_{4}$ (where BEDT-TSF is Bis(ethylenedithio)tetraselenafulvalene, C$_{10}$S$_{4}$Se$_{4}$H$_{8}$),
J. Phys. Soc. Jpn. {\bf 69,} 2759 (2000).

\bibitem{SpinFlopSmFeAsO}
S. Weyeneth, P. J. W. Moll, R. Puzniak, K. Ninios, F. F. Balakirev, R. D. McDonald, H. B. Chan, N. D. Zhigadlo, S. Katrych, Z. Bukowski, J. Karpinski, H. Keller, B. Batlogg, and L. Balicas,
Rearrangement of the antiferromagnetic ordering at high magnetic fields in SmFeAsO and SmFeAsO$_{0.9}$F$_{0.1}$ single crystals,
Phys. Rev. B {\bf 83,} 134503 (2011).

\bibitem{SpinFlopCaFePtAs}
M. D. Watson, A. McCollam, S. F. Blake, D. Vignolles, L. Drigo, I. I. Mazin, D. Guterding, H. O. Jeschke, R. Valent\'{i}, N. Ni, R. Cava, and A. I. Coldea,
Field-induced magnetic transitions in Ca$_{10}$(Pt$_3$As$_8$)((Fe$_{1-x}$Pt$_x$)$_2$As$_2$)$_5$ compounds,
Phys. Rev. B {\bf 89,} 205136 (2014)


\bibitem{PhaseDiagram}
J Bernhard, B Lebech and O. Beckman,
Magnetic phase diagram of hexagonal FeGe determined by neutron diffraction,
J. Phys. F {\bf 18,} 539 (1988).

\bibitem{STMYajunYan}
Z. Chen, X. Wu, R. Yin, J. Zhang, S. Wang, Y, Li, M. Li, A. Wang, Y. Wang, Y.-J. Yan, and D.-L. Feng,
Charge density wave with strong quantum phase fluctuations in Kagome magnet FeGe,
arXiv: 2302.04490 (2023).


\bibitem{CDWLinearMR}
Y. Feng, Y. Wang, D. M. Silevitch, J.-Q. Yan, R. Kobayashi, M. Hedo, T. Nakama, Y. \={O}nuki, A. V. Suslov, B. Mihaila, P. B. Littlewood, and T. F. Rosenbaum,
Linear magnetoresistance in the low-field limit in density-wave materials,
Proc. Natl. Acad. Sci. USA {\bf 116,} 11201 (2019).

\bibitem{VASP}
J. Hafner,
Ab-initio simulations of materials using VASP: Density-functional theory and beyond,
J. Comput. Chem. {\bf 29,} 2044 (2008).

\bibitem{Perdew}
J. P. Perdew, K. Burke and M. Ernzerhof,
Generalized gradient approximation made simple,
Phys. Rev. Lett. {\bf 77,} 3865 (1996).


\bibitem{PristineStructure}
M. Richardson,
The Partial Equilibrium Diagram of the Fe-Ge System in the Range 40-72 at. \% Ge, and the Crystallisation of some Iron Germanides by Chemical Transport Reactions,
Acta Chem. Scan {\bf 21,} 2305 (1967).

\bibitem{Wannier90}
A. A. Mostofi, J. R. Yates, Y.-S. Lee, I. Souza, D. Vanderbilt, and N. Marzari,
wannier90: A tool for obtaining maximally-localised Wannier functions,
Comput. Phys. Commun. {\bf 178,} 685 (2008).

\bibitem{SKEAF}
P. M. C. Rourke, S. R. Julian,
Numerical extraction of de Haas--van Alphen frequencies from calculated band energies,
Comput. Phys. Commun. {\bf 183,} 324 (2012).

\bibitem{JiangARPES}
Z. Zhao {\it et al.},
Photoemission Evidence of a Novel Charge Order in Kagome Metal FeGe,
arXiv: 2308.08336 (2023).







\bibitem{Shoenberg}
D. Shoenberg, {\it Magnetic Oscillations in Metals}, (Cambridge University Press, Cambridge, England, 1984).

\bibitem{footnote}
A local spin density approximation calculation \cite{WanNesting} provides a density of states $\sim$3 states/eV at the Fermi level; this value roughly matches the result of a specific heat measurement (Aifeng Wang, unpublished). Considering a Hall carrier density of $\sim$2.5-3.1$\times$10$^{22}$ cm$^{-3}$ [Fig.\,S3(c), \cite{SM}], the effective mass ratio $m^*$/$m_e$ is estimated to be larger than 8. The lack of evidence for such an renormalization factor in dHvA study implies that the quasiparticle effective masses vary drastically between different bands, whereas our measurements only capture orbits with the lightest masses.

\bibitem{meanfreepath}
$R_D$ = $\exp[-14.69(m^*/m_e)(T_D/B)]$; the ``Dingle temperature" $T_D$ = $\hbar(2\pi k_B \tau_{QP})^{-1}$ ($\tau_{QP}$ is the quasiparticle scattering time, $k_B$ is the Boltzmann constant) \cite{Shoenberg}. Using a single-component approximation, we obtain $T_D$ = (10$\pm$3)\,K for branch $\alpha$ (Fig.\,S4 in \cite{SM}), which converts to $\tau_{QP}$ = (1.2$\pm$0.3)$\times$10$^{-13}$\,s. The quasiparticle mean free path is given by $l_{QP} = v_F\tau_{QP}$. Here, the Fermi velocity $v_F = \hbar k_F/m^*$ can be estimated from the experimentally determined Fermi vector $k_F$ (= $[2e F/\hbar]^{1/2}$) and cyclotron mass $m^*$. For dHvA branch $\alpha$ ($F$ =445\,T, $m^*$ = 0.20 $m_e$), $v_F$ is calculated to be 6.7$\times$10$^5$\,m/s.

\bibitem{WilsonTaSe2}
J. A. Wilson,
Charge-density waves in the 2H-TaSe$_2$ family: Action on the Fermi surface,
Phys. Rev. B {\bf 15,} 5748 (1977).

\bibitem{Harrison_AFM}
N. Harrison, R. D. McDonald, and J. Singleton,
Cuprate Fermi orbits and Fermi arcs: the effect of short-range antiferromagnetic order,
Phys. Rev. Lett. {\bf 99,} 206406 (2007).

\bibitem{ShortRangeCDW}
Y. Gannot, B. J. Ramshaw, and S. A. Kivelson,
Fermi surface reconstruction by a charge density wave with finite correlation length,
Phys. Rev. B {\bf 100.} 045128 (2019).

\bibitem{CDWdamping}
Accordinig to \cite{ShortRangeCDW}, $R_{\rm CDW} = \exp[-4(r_c/\xi)]$ for the 2$\times$2 charge order in FeGe. Hence, with $r_c = \hbar k_F/eB \approx$ 26\,nm for $F_{\alpha}$ at $\mu_0H \simeq$30\,T, a small $\xi \lesssim 4$\,nm reported for samples in batch ``C" \cite{STMYajunYan} damps the dHvA amplitudes by a factor of 10$^{12}$, precluding any experimental observation.

\bibitem{Tl2201RIXS}
C. C. Tam, M. Zhu, J. Ayres, K. Kummer, F. Yakhou-Harris, J. R. Cooper, A. Carrington, and S. M. Hayden,
Charge density waves and Fermi surface reconstruction in the clean overdoped cuprate superconductor Tl$_2$Ba$_2$CuO$_{6+\delta}$,
Nat. Commun. {\bf 13,} 570 (2022).


\bibitem{YBCO2008Nature}
S. E. Sebastian, N. Harrison, E. Palm, T. P. Murphy, C. H. Mielke, R. Liang, D. A. Bonn, W. N. Hardy, and G. G. Lonzarich,
A multi-component Fermi surface in the vortex state of an underdoped high-$T_c$ superconductor,
Nature (London) {\bf 454,} 200 (2008).

\bibitem{YBCOdHvA}
C. Jaudet, D. Vignolles, A. Audouard, J. Levallois, D. LeBoeuf, N. Doiron-Leyraud, B. Vignolle, M. Nardone, A. Zitouni, R. Liang, D. A. Bonn, W. N. Hardy, L. Taillefer, and C. Proust,
de Haas-van Alphen Oscillations in the Underdoped High-Temperature Superconductor YBa$_2$Cu$_3$O$_{6.5}$,
Phys. Rev. Lett. {\bf 100,} 187005 (2008).

\bibitem{YBCO_XRS}
J. Chang, E. Blackburn, O. Ivashko, A. T. Holmes, N. B. Christensen, M. H\"{u}cker, R. Liang, D. A. Bonn, W. N. Hardy, U. R\"{u}tt, M. v. Zimmermann, E. M. Forgan, and S. M. Hayden,
Magnetic field controlled charge density wave coupling in underdoped YBa$_2$Cu$_3$O$_{6+x}$,
Nat. Commun. {\bf 7,} 11494 (2016).

\bibitem{Hg1201Barisic}
N. Bari\u{s}i\'{c}, S. Badoux, M. K. Chan, C. Dorow, W. Tabis, B. Vignolle, G. Yu, J. B\'{e}ard, X. Zhao, C. Proust, and M. Greven,
Universal quantum oscillations in the underdoped cuprate superconductors,
Nat. Phys. {\bf 9,} 761 (2013).

\bibitem{Hg1201MKChan}
M. K. Chan, N. Harrison, R. D. McDonald, B.J. Ramshaw, K. A. Modic, N. Bari\u{s}i\'{c}, and M. Greven,
Single reconstructed Fermi surface pocket in an underdoped single-layer cuprate superconductor,
Nat. Commun. {\bf 7,} 12244 (2016).

\bibitem{cuprates}
For instance, YBa$_2$Cu$_3$O$_{6+x}$ \cite{YBCO2008Nature,YBCOdHvA,YBCO_XRS} and HgBa$_2$CuO$_{4+\delta}$ \cite{Hg1201Barisic,Hg1201MKChan} with ($\xi$, $l_{QP}$) = ($\sim$ 30\,nm, 16-17\,nm) and (2\,nm, 8.5\,nm), respectively.

\bibitem{LengthEstimation}
Samples from batch ``A" were grown following the method reported in Ref.\,\cite{FeGeCDW_Neutron}, wherein a short $\xi$ of 3.3-4.5\,nm is determined by neutron scattering. Nevertheless, the presence of dHvA signals suggests that the ratio $r_c/\xi$ should not be very large \cite{ShortRangeCDW,CDWdamping}. Considering $l_{QP} \sim$ 60-100\,nm (Figure S4, \cite{SM}) and $r_c \approx$ 26\,nm (at 30\,T) for $F_{\alpha}$, we speculate $\xi \sim$ 20-40\,nm for our batch ``A".

\bibitem{CsV3Sb5HighF}
C. Broyles, D. Graf, H. Yang, X. Dong, H. Gao, and S. Ran,
Effect of the interlayer ordering on the Fermi surface of Kagome superconductor CsV$_3$Sb$_5$ revealed by quantum oscillations,
Phys. Rev. Lett. {\bf 129,} 157001 (2022).

\bibitem{CsV3Sb5MB}
R. Chapai, M. Leroux, V. Oliviero, D. Vignolles, N. Bruyant, M. P. Smylie, D. Y. Chung, M. G. Kanatzidis, W.-K. Kwok, J. F. Mitchell, and U. Welp,
Magnetic Breakdown and Topology in the Kagome Superconductor CsV$_3$Sb$_5$ under High Magnetic Field,
Phys. Rev. Lett. {\bf 130,} 126401 (2023).

\bibitem{Ta4Pd3Te16}
T. Helm, F. Flicker, R. Kealhofer, P. J. W. Moll, I. M. Hayes, N. P. Breznay, Z. Li, S. G. Louie, Q. R. Zhang, L. Balicas, J. E. Moore, and J. G. Analytis,
Thermodynamic anomaly above the superconducting critical temperature in the quasi-one-dimensional superconductor Ta$_4$Pd$_3$Te$_{16}$,
Phys. Rev. B {\bf 95,} 075121 (2017).
















\end{thebibliography}
\end{document}